\documentclass[twocolumn,showpacs,amsfonts,aps,prc,floatfix,superscriptaddress,nofootinbib,groupedaddress]{revtex4-1}
\usepackage{CJK}
\usepackage{multirow}
\usepackage{amssymb}
\usepackage{amsmath}
\usepackage{mciteplus}
\usepackage{hyperref}
\usepackage{graphicx}
\usepackage{dcolumn}
\usepackage{bm}
\usepackage{lineno}
\usepackage{float}
\usepackage{subfloat}
\usepackage{color}

\usepackage{epstopdf}
\usepackage[normalem]{ulem}

\usepackage{color}
\definecolor{dgreen}{cmyk}{1.,0.,1.,0.1}        
\definecolor{orange}{cmyk}{0.,0.353,1.,0.}    

 \newif\ifpdf
 \ifx\pdfoutput\undefined
 \pdffalse
 \else
 \pdfoutput=1
 \pdftrue
 \fi
 \ifpdf
 \usepackage{graphicx}
 \usepackage{epstopdf}
 \else
 \usepackage{graphicx}
 \fi

\newcommand{\pt}    {p_{\rm T}}

\begin{document}
\begin{CJK*}{GB}{}

\title{Correlations of flow harmonics in 2.76A TeV Pb--Pb collisions}

\author{Xiangrong Zhu$^{1,2,3}$, You Zhou$^{4}$, Haojie Xu$^{1,2}$, Huichao Song$^{1,2,5}$}
\email{You Zhou: you.zhou@cern.ch; \\
Huichao Song: Huichaosong@pku.edu.cn}
\affiliation{$^{1}$Department of Physics and State Key Laboratory of Nuclear Physics and Technology, Peking University, Beijing 100871, China}
\affiliation{$^{2}$Collaborative Innovation Center of Quantum Matter, Beijing 100871, China}
\affiliation{$^3$School of Science, Huzhou University, Huzhou 313000, China}
\affiliation{$^{4}$Niels Bohr Institute, University of Copenhagen, Blegdamsvej 17, 2100 Copenhagen, Denmark}
\affiliation{$^{5}$Center for High Energy Physics, Peking University, Beijing 100871, China}

\date{\today}

\begin{abstract}
Using the event-by-event viscous hydrodynamics {\tt VISH2+1} with {\tt MC-Glauber}, {\tt MC-KLN}, and {\tt AMPT} initial conditions, we investigate the correlations of flow harmonics, including the symmetric cumulants $SC^{v}(m, n)$, the normalized symmetric cumulants $NSC(m, n)$, and the Pearson correlation coefficients $C(v_{m}^{2}, v_{n}^{2})$ in 2.76A TeV Pb--Pb collisions. We find $SC^{v}(m, n)$ is sensitive to both initial conditions and the specific shear viscosity $\eta/s$. A comparison with the recent ALICE data show that our hydrodynamic calculations can qualitatively describe the data of $SC^{v}(3, 2)$ and $SC^{v}(4, 2)$ for various initial conditions, which demonstrate that $v_2$, $v_4$ are correlated and $v_2$, $v_3$ are anti-correlated. Meanwhile, the predicted symmetric cumulants $SC^{v}(5, 2)$, $SC^{v}(5, 3)$, and $SC^{v}(4, 3)$ reveal that $v_2$ and $v_5$, $v_3$ and $v_5$ are correlated, $v_3$ and $v_4$ are anti-correlated in most centrality classes. We also find $NSC^{v}(3, 2)$ and $C(v_{3}^{2}, v_{2}^{2})$, which are insensitive to $\eta/s$, are mainly determined by corresponding $NSC^{\varepsilon}(3, 2)$ and $C(\varepsilon_{3}^{2}, \varepsilon_{2}^{2})$ correlators from the initial state. In contrast, other $NSC^{v}(m, n)$ and $C(v_{m}^{2}, v_{n}^{2})$ correlators are influenced by both initial conditions and $\eta/s$, which illustrates the non-linear mode couplings in higher flow harmonics with $n \geq 4$.

\end{abstract}

\pacs{25.75.-q, 12.38.Mh, 25.75.Ld, 24.10.Nz}
\maketitle

\end{CJK*}

\section{INTRODUCTION}
The main goals of ultrarelativistic heavy-ion collisions at the BNL Relativistic Heavy Ion Collider (RHIC) and the CERN Large Hadron Collider (LHC) are to produce the strongly interacting Quark-Gluon Plasma (QGP), a deconfined state of quarks and gluons, and to explore its properties~\cite{Rev-Arsene:2004fa,Gyulassy:2004vg,Muller:2006ee,Muller:2012zq}. The azimuthal anisotropy of produced hadrons is one of the important observables to probe the prosperities of the QGP~\cite{Ollitrault:1992bk,Voloshin:2008dg}. It could be characterized by an expansion of the single-particle azimuthal distributions $P(\varphi)$:
\begin{equation}
P(\varphi) = \frac{1}{2\pi} \sum_{n=-\infty}^{+\infty} {\overrightarrow{V_{n}} \, e^{-in\varphi} }
\end{equation}
where $\varphi$ is the azimuthal angle of the emitted particles, $\overrightarrow{V_{n}}$ is the $n$-th order flow-vector, defined as $\overrightarrow{V_{n}} = v_{n}\,e^{in\Psi_{n}}$. Its magnitude $v_{n}$ is the $n$-th order anisotropic flow harmonics and its orientation $\Psi_{n}$ is the symmetry plane angle. The anisotropic flow harmonics $v_{n}$ have been studied in great details by many groups (For a recent review, please see~\cite{ Luzum:2013yya,Heinz:2013th, Gale:2013da}). The observation of elliptic flow and higher order flow harmonics at RHIC and the LHC, together with the successful descriptions from hydrodynamics and hybrid models, demonstrates that the QGP fireball fluctuates event-by-event and behaves like a nearly perfect liquid with a very small specific shear viscosity~\cite{ALICE:2011ab,Chatrchyan:2013kba, Luzum:2013yya,Heinz:2013th,Gale:2013da,Qiu:2011iv,Petersen:2010cw,Holopainen:2010gz,Song:2010mg,Song:2013qma,Song:2012ua,Schenke:2010rr, Gale:2012rq,Xu:2016hmp}. Besides the flow harmonics $v_n$, additional information for the initial state fluctuations can be obtained by studying the correlations between different order flow-vectors $\overrightarrow{V_{m}}$ and $\overrightarrow{V_{n}}$. Initially, the study of the correlations between the orientations of different flow-vector was investigated in the observable of $v_{2n/\Psi_{n}}$~\cite{Andronic:2000cx, Chung:2001qr, Adams:2003zg}. Recently, a full systematic study of so-called ``event-plane correlations'' was carried by the ATLAS Collaboration~\cite{Aad:2014fla}. The corresponding hydrodynamic simulations and related theoretical investigations suggest that these new correlations open a new window to probe the details of initial state fluctuations and the transport properties of the QGP~\cite{Qiu:2012uy,Teaney:2012gu, Jia:2012ju,Jia:2012ma, Niemi:2015qia}.

In addition, the correlations between different order flow harmonics can be used to further investigate the details of initial-state fluctuations and the hydrodynamic response~\cite{Bilandzic:2013kga, Bhalerao:2014xra,Zhou:2015eya,ALICE:2016kpq,Niemi:2012aj,Giacalone:2016afq,Qian:2016pau}. They also reveal whether different order flow harmonics $v_{m}$ and $v_{n}$ are correlated, anti-correlated or uncorrelated. On the experimental side, it is crucial to find an observable that measures the flow harmonics correlations without contributions from the symmetry plane correlations. The first experimental attempt was made by the ATLAS Collaboration in~\cite{Aad:2015lwa}. They investigated the $v_{m}$ and $v_{n}$ correlations for events within a given narrow centrality class using an Event-Shape Engineering (ESE), a technique to select events according to the magnitude of reduced flow vector $\overrightarrow{q_{n}}$~\cite{Schukraft:2012ah}. It was observed that, for events within the same centrality class, $v_{2}$ is anti-correlated with $v_{3}$ and correlated with $v_{4}$~\cite{Aad:2015lwa}. However, this measurement was based on the 2-particle correlations, which might be largely contaminated by non-flow effects. Their method also requires sub-dividing such calculations and modeling resolutions associated with ESE due to finite event-wise multiplicities. Considering these constraints, this approach can not be easily utilized in the hydrodynamic simulations with fluctuating initial conditions, as reported in~\cite{Qian:2016pau}.

In Ref.~\cite{Niemi:2012aj}, it is suggested to study the correlations between $v_{m}$ and $v_{n}$ through the linear correlation function $c(v_{m},v_{n})$. However, this observable can not be easily accessed in experiments, in which the measurements rely on two- and multi-particle correlations techniques. Later on, a new observable called Symmetric Cumulants $SC^{v}(m, n)$, which can be measured by the multi-particle cumulant method~\cite{Bilandzic:2013kga}, is proposed to study the correlations between different flow harmonics. It is supposed to be insensitive to the non-flow effects and is free of symmetry plane correlations by design~\cite{Bilandzic:2013kga}. Recently, $SC^{v}(4, 2)$ and $SC^{v}(3, 2)$ are measured by the ALICE Collaboration~\cite{ALICE:2016kpq}. Positive values of $SC^{v}(4, 2)$ and negative values of $SC^{v}(3, 2)$ are observed at various centrality bins~\cite{ALICE:2016kpq}, which suggests that $v_{2}$ and $v_{4}$ are correlated and $v_{2}$ and $v_{3}$ are anti-correlated. Meanwhile, the {\tt HIJING} model simulations, which do not include the collective expansion, show that although the non-flow effects lead to non-zero values for both the 2-particle correlations $\left< v_{n}^{2} \right>$, and the 4-particle correlations $\left< v_{m}^{2}v_{n}^{2} \right>$, the 4-particle cumulants $SC^{v}(4, 2)$ and $SC^{v}(3, 2)$ from {\tt HIJING} are consistent with zero. This suggests that $SC^{v}(m, n)$ is an ideal observable that evaluates the correlations between flow harmonics, and is insensitive to non-flow effects.

In this paper, we will investigate the correlations of flow harmonics with event-by-event viscous hydrodynamics {\tt VISH2+1}~\cite{Song:2007fn,Shen:2014vra}. To study the influences from initial conditions and explore the general properties of the final state correlations, we implement three different initial conditions, namely, {\tt MC-Glauber}, {\tt MC-KLN}, and {\tt AMPT} initial conditions. We will compare our calculated Symmetric Cumulants $SC^{v}(4, 2)$ and $SC^{v}(3, 2)$  with the ALICE data and predict other Symmetric Cumulants. We will also study the normalized Symmetric Cumulants $NSC(m, n)$, the Pearson correlation coefficients $C(v_{m}^{2}, v_{n}^{2})$ (definitions seen Section~\ref{sec:setup}) and explore their sensitivities to initial conditions and the specific shear viscosity.

The paper is organized as follows. Section~\ref{sec:setup} introduces the {\tt VISH2+1} hydrodynamic model, the setup of calculations and the methodology to calculate the correlations of flow harmonics. Section~\ref{sec:results} presents the results and discussions for the correlations of flow harmonics. Section~\ref{sec:summary} provides a brief summary of this paper.

\section{The model and the setup of the calculations\label{sec:setup}}
In this paper, we implement event-by-event viscous hydrodynamics {\tt VISH2+1}~\cite{Song:2007fn,Shen:2014vra} to study the correlations of flow harmonics in 2.76A TeV Pb--Pb collisions. {\tt VISH2+1} is a (2+1)-d viscous hydrodynamic code to solve the transport equations of the energy momentum tensor and the time evolution equations of the shear stress tensor and bulk pressure based on the Israel-Stewart formalism, which simulates the viscous fluid expansion of the hot QCD matter with longitudinal boost-invariance~\cite{Song:2007fn}. Around 2011, it was updated to an event-by-event simulation version to further study the initial state fluctuations and final state correlations~\cite{Qiu:2011hf}. The equation of state, initial and decoupling  conditions, as well as the transport coefficients are additional inputs of the {\tt VISH2+1} code. Generally, {\tt VISH2+1} implements the state of the art equation of state EoS-s95p-PCE, which could account for the partially chemical equilibrium effects during the hadronic evolution~\cite{Huovinen:2009yb,Shen:2010uy}.
Along a decoupling hyper-surface, which is generally defined by a constant temperature $T_{dec}$, the hydrodynamic information is converted to the final hadron distributions though the Cooper-Fryer formula~\cite{Cooper:1974mv}. Compared with the hybrid model approach that connects viscous hydrodynamics with a hadron cascade model at a switching temperature near $T_c$ (i.e. {\tt iEBE-VISHNU}~\cite{Shen:2014vra}), the {\tt VISH2+1} simulations implemented in this paper describe both the QGP fluid and the highly dissipative and even off-equilibrium late hadronic stage with fluid-dynamics. With well tuned transport coefficients, decoupling temperature $T_{dec}$ and other related parameters, together with some well-chosen initial conditions (like {\tt AMPT}~\cite{Xu:2016hmp,Bhalerao:2015iya,Pang:2012he} and {\tt T\raisebox{-.5ex}{R}ENTo}~\cite{Moreland:2014oya}, etc.), it could fit many related soft hadron data, such as the $\pt$ spectra and different flow harmonics at RHIC and the LHC~\cite{Qiu:2011hf, Shen:2010uy, Shen:2011eg, Bhalerao:2015iya}. Note that this paper does not aim to precisely fit the flow data to extract some information of the hot QCD matter, but focuses on exploring the general properties of the correlations between different flow harmonics. We thus implement the computational efficient {\tt VISH2+1} code and leave the more sophisticated but also calculation-time consuming hybrid model simulations to the future study.

To investigate the dependence of flow harmonics correlations on the initial state, we implement three different initial conditions, namely, {\tt MC-Glauber}, {\tt MC-KLN} and {\tt AMPT} initial conditions in the following hydrodynamic calculations. Traditionally, the Glauber model constructs the initial entropy density of the QGP fireball from a mixture of the wounded nucleon and binary collision density profiles~\cite{Kolb:2000sd}, and the {\tt KLN} model assumes the initial entropy density is proportional to the initial gluon density calculated from the corresponding $k_T$ factorization formula~\cite{Kharzeev:2000ph}. In the Monte-Carlo versions ({\tt MC-Glauber} and {\tt MC-KLN})~\cite{Miller:2007ri,Drescher:2006ca,Hirano:2009ah}, additional initial state fluctuations are introduced through the position fluctuations of individual nucleons inside the colliding nuclei. For the {\tt AMPT} initial conditions~\cite{Bhalerao:2015iya,Pang:2012he,Xu:2016hmp}, the fluctuating energy density profiles are constructed from the energy decompositions of individual partons, which fluctuate in both momentum and position space. Compared with the {\tt{MC-Glauber}} and {\tt{MC-KLN}} initial conditions, the additional Gaussian smearing parameter in the {\tt AMPT} initial conditions makes the typical initial fluctuation scales changeable, which helps to achieve better hydrodynamic descriptions of the anisotropic flow data.

Considering the conversion from initial entropy to final multiplicity of all charged hadrons, the centrality is determinated via the distribution of total entropies of the fluctuating initial profiles. In order to explore the shear viscosity dependence of flow harmonic correlations, we choose two values of $\eta/s$, 0.08 and 0.20, for {\tt MC-Glauber} and {\tt MC-KLN} initial conditions, and 0.08 and 0.16 for {\tt AMPT} initial conditions~\footnote{We have noticed, in peripheral Pb--Pb collisions, {\tt VISH2+1} simulations with {\tt AMPT} initial conditions and $\eta/s=0.20$ are roughly out of the validity regime of hydrodynamics due to large viscous corrections~\cite{Shen:2014vra,Oliinychenko:2015lva,Xu-Song}. We thus choose a smaller value of the specific shear viscosity $\eta/s=0.16$ for the corresponding comparison runs.}. For each initial condition and $\eta/s$, the normalization factors of the initial entropy density profiles and the hydrodynamic starting time $\tau_0$ are respectively tuned to fit the multiplicity and $\pt$ spectra of all charged hadrons in the most central Pb--Pb collisions~\cite{Xu:2016hmp}. Following~\cite{Xu:2016hmp,Shen:2011eg}, the decoupling temperature $T_{dec}$ is set to 120 MeV, which could roughly describe the slope of the $\pt$ spectra of protons in the most central collisions. To simplify the theoretical investigations, we set the bulk viscosity, net baryon density and the heat conductivity to zero in the following calculations.

After the hydrodynamic evolution and thermal freeze-out, the  anisotropic flow coefficients $v_{n}$ and its corresponding symmetry plane angle $\Psi_{n}$ can be calculated as~\cite{Qiu:2011iv,Shen:2014vra}:
\begin{equation}
  v_n\,e^{i n \Psi_n} =\frac{\int \pt\,d\pt\,d\varphi\, e^{i n\varphi}\,\frac{dN_\mathrm{ch}^{3}}{\pt d\pt\,d\varphi\, d\eta}}
  {\int \pt\, d\pt\,d\varphi\,\frac{dN_\mathrm{ch}^{3}}{\pt\, d\pt\,d\varphi\,d\eta}}
  \label{Eq:vnPsin}
\end{equation}
where $\varphi$ is the azimuthal angle of the emitted particles, $n$ is the order of the flow harmonics.

With $v_{n}$ obtained from the above equation, one can calculate the Symmetric Cumulants, $SC^{v}(m, n)$ defined as the following~\cite{ALICE:2016kpq}:
\begin{equation}
SC^{v}(m, n)= \left< v_{m}^{2} \, v_{n}^{2} \right> - \left< v_{m}^{2} \right> \left< v_{n}^{2} \right>.
\end{equation}

To further evaluate the correlations of flow harmonics, one could define the normalized Symmetric Cumulants:
\begin{equation}
NSC^{v}(m, n) = \frac{SC^{v}(m, n)}{\langle v_{m}^{2}\rangle \langle v_{n}^{2}\rangle}=\frac{\langle v_{m}^{2}v_{n}^{2}\rangle -  \langle v_{m}^{2}\rangle \langle v_{n}^{2}\rangle}{\langle v_{m}^{2}\rangle \langle v_{n}^{2}\rangle}
\label{eq:NSCvmvn}
\end{equation}
Compared with $SC^{v}(m, n)$, $NSC^{v}(m, n)$ reflects the relative correlation between $v_{m}$ and $v_{n}$, which is expected to be insensitive to the magnitudes of $v_{m}$ and $v_{n}$.

Alternatively, the correlations between flow harmonics $v_{m}$ and $v_{n}$ can be investigated via the Pearson correlation coefficient, which has been widely used to evaluate the degree of linear dependence between two variables~\cite{Niemi:2012aj,PCC}. The Pearson correlation coefficient is defined as:
\begin{eqnarray}
C(v_{m}^{2}, v_{n}^{2}) &=& \rho_{v_{m}^{2},v_{n}^{2}} =\frac{\langle(v_{m}^{2}-\langle v_{m}^{2}\rangle)(v_{n}^{2}-\langle v_{n}^{2}\rangle)\rangle}{\sigma_{v_{m}^{2}}\sigma_{v_{n}^{2}}} \nonumber \\
                          &=& \frac{\langle v_{m}^{2}v_{n}^{2}\rangle-\langle v_{m}^{2}\rangle\langle v_{n}^{2}\rangle}{\sqrt{\langle v_{m}^{4}\rangle-\langle v_{m}^{2}\rangle^{2}}\sqrt{\langle v_{n}^{4}\rangle-\langle v_{n}^{2}\rangle^{2}}}.
 \label{eq:SCCRvmvn}
\end{eqnarray}
where $\sigma_{v_{m}}$ stands for the standard deviation of $v_{m}$ distributions(In the flow language, it is also the flow fluctuations of $v_{m}$). Generally speaking, $C(v_{m}^{2}, v_{n}^{2}) =$ 1 or -1 means that the variables $v_{m}$ and $v_{n}$ are total linearly correlated or anti-correlated, $C(v_{m}^{2}, v_{n}^{2}) =$ 0 means $v_{m}$ and $v_{n}$ are uncorrelated.

Correspondingly, one could also investigate the correlations between different eccentricity coefficients. For a fluctuating initial profile, the eccentricity coefficients $\varepsilon_{n}$ and the initial symmetry plane (participant plane) angle $\Phi_{n}$ are defined as~\cite{Qiu:2011iv,Shen:2014vra}:
\begin{equation}
\label{eq:eccCalculation}
 \varepsilon_{n}e^{in\Phi_{n}}=-\frac{\int r\, dr\, d\varphi\, r^n\, e^{in\phi}\, e(r, \varphi)}{\int r\, dr\, d\varphi\, r^n\, e(r, \varphi)},
\end{equation}
where $e(r, \varphi)$ is the initial energy density in the transverse plane, $\varphi$ is azimuthal angle and $n$ is the order of the coefficient. Analogous to Eqs.~(3-5), we propose the corresponding correlators $SC^{\varepsilon}(m, n)$, $NSC^{\varepsilon}(m, n)$, and $C(\varepsilon_{m}^{2}, \varepsilon_{n}^{2})$ for the initial state, which are defined as following:
\begin{equation}
\label{eq:SCemen}
SC^{\varepsilon}(m, n)=\langle\varepsilon_{m}^{2}\varepsilon_{n}^{2}\rangle-\langle\varepsilon_{m}^{2}\rangle\langle\varepsilon_{n}^{2}\rangle,
\end{equation}
\begin{eqnarray}
NSC^{\varepsilon}(m, n)&=&\frac{SC^{\varepsilon}(m, n)}{\langle \varepsilon_{m}^{2}\rangle \langle \varepsilon_{n}^{2}\rangle}=\frac{\langle \varepsilon_{m}^{2}\varepsilon_{n}^{2}\rangle - \langle \varepsilon_{m}^{2}\rangle \langle \varepsilon_{n}^{2}\rangle}{\langle \varepsilon_{m}^{2}\rangle \langle \varepsilon_{n}^{2}\rangle},
\label{eq:NSCemen}
\end{eqnarray}
and
\begin{eqnarray}
C(\varepsilon_{m}^{2}, \varepsilon_{n}^{2}) &=& \rho_{\varepsilon_{m}^{2},\varepsilon_{n}^{2}} =\frac{\langle(\varepsilon_{m}^{2}-\langle \varepsilon_{m}^{2}\rangle)(\varepsilon_{n}^{2}-\langle \varepsilon_{n}^{2}\rangle)\rangle}{\sigma_{\varepsilon_{m}^{2}}\sigma_{\varepsilon_{n}^{2}}} \nonumber \\
                          &=&\frac{\langle \varepsilon_{m}^{2}\varepsilon_{n}^{2}\rangle-\langle \varepsilon_{m}^{2}\rangle\langle \varepsilon_{n}^{2}\rangle}{\sqrt{\langle \varepsilon_{m}^{4}\rangle-\langle \varepsilon_{m}^{2}\rangle^{2}}\sqrt{\langle \varepsilon_{n}^{4}\rangle-\langle \varepsilon_{n}^{2}\rangle^{2}}}
\label{eq:SCCRemen}
\end{eqnarray}

\begin{figure*}[htpb]
\centering
 \includegraphics[width=0.8\linewidth]{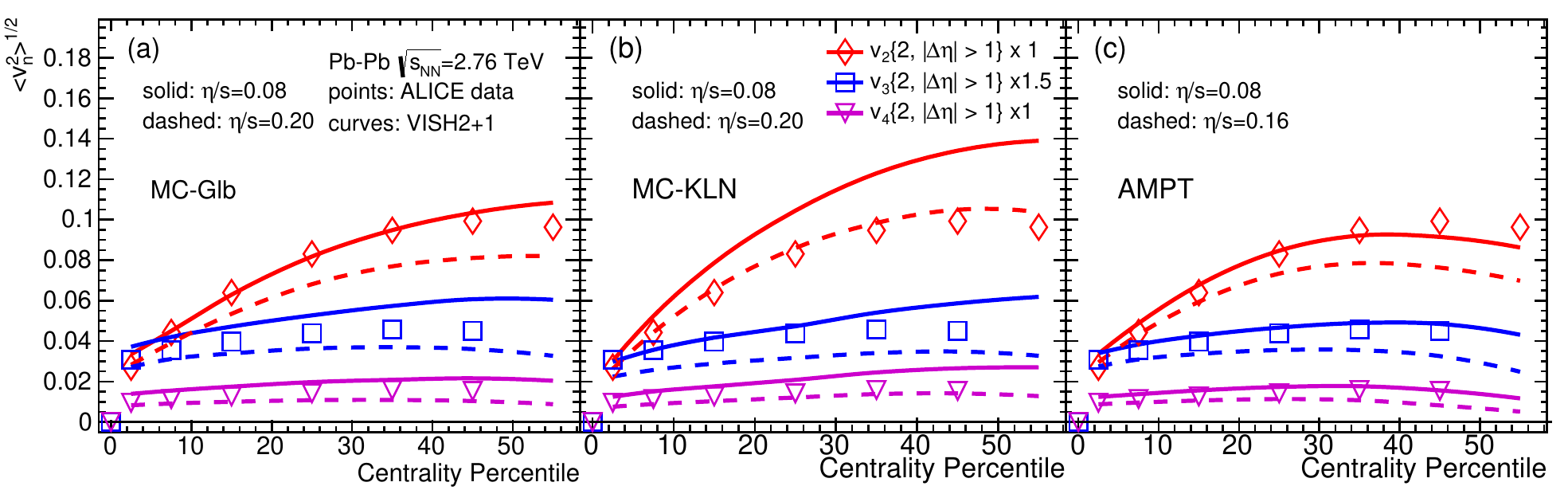}
  \caption{(Color online) Integrated flow $v_{2}$ (red curves, $\times 1$), $v_{3}$ (blue curves, $\times 1.5$), and $v_{4}$ (magenta curves, $\times 1$) of all charged hadrons in 2.76 A TeV Pb--Pb collisions, calculated from {\tt VISH2+1} with {\tt MC-Glauber} (left), {\tt MC-KLN} (middle), and {\tt AMPT} (right) initial conditions, together with a comparison to the ALICE data~\cite{ALICE:2011ab}.
 \label{fig:v2v3v4}}
\end{figure*}

\begin{figure*}[ht]
\centering
 \includegraphics[width=0.8\linewidth]{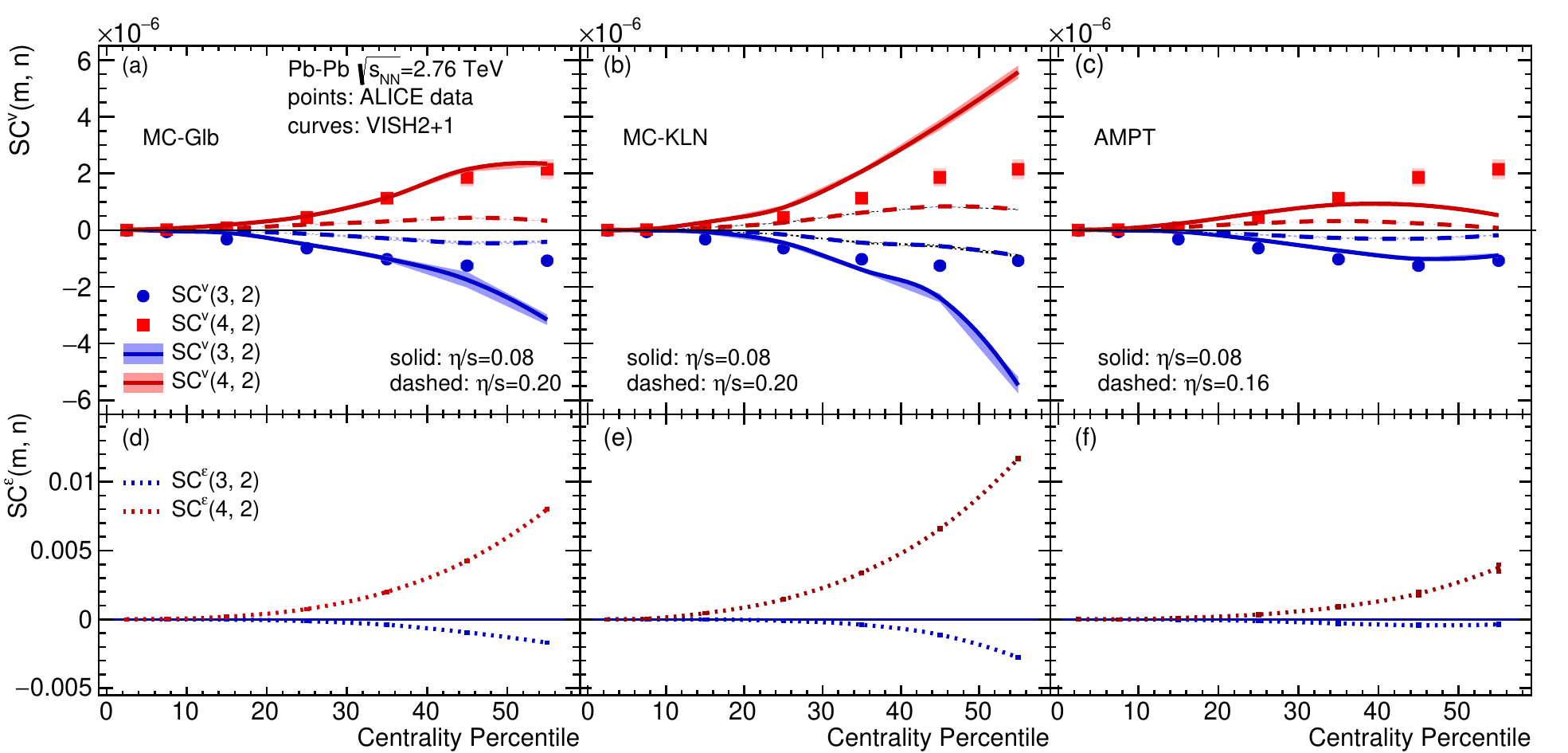}
 \caption{(Color online) Top panels: Symmetric cumulants $SC^{v}(3, 2)$ and $SC^{v}(4, 2)$ in 2.76 A TeV Pb--Pb collisions, calculated from {\tt VISH2+1} with {\tt MC-Glauber} (a), {\tt MC-KLN} (b), and {\tt AMPT} (c) initial conditions and with different $\eta/s$. The measured $SC^{v}(3, 2)$ and $SC^{v}(4, 2)$ from the ALICE Collaboration are also presented here, which are taken from~\cite{ALICE:2016kpq}. Bottom panels: Symmetric cumulants of the initial eccentricity coefficients $SC^{\varepsilon}(3, 2)$ and $SC^{\varepsilon}(4, 2)$ for these three initial conditions.
 \label{fig:scvmvn1}}
\end{figure*}

\begin{figure*}[htpb]
\centering
 \includegraphics[width=0.8\linewidth]{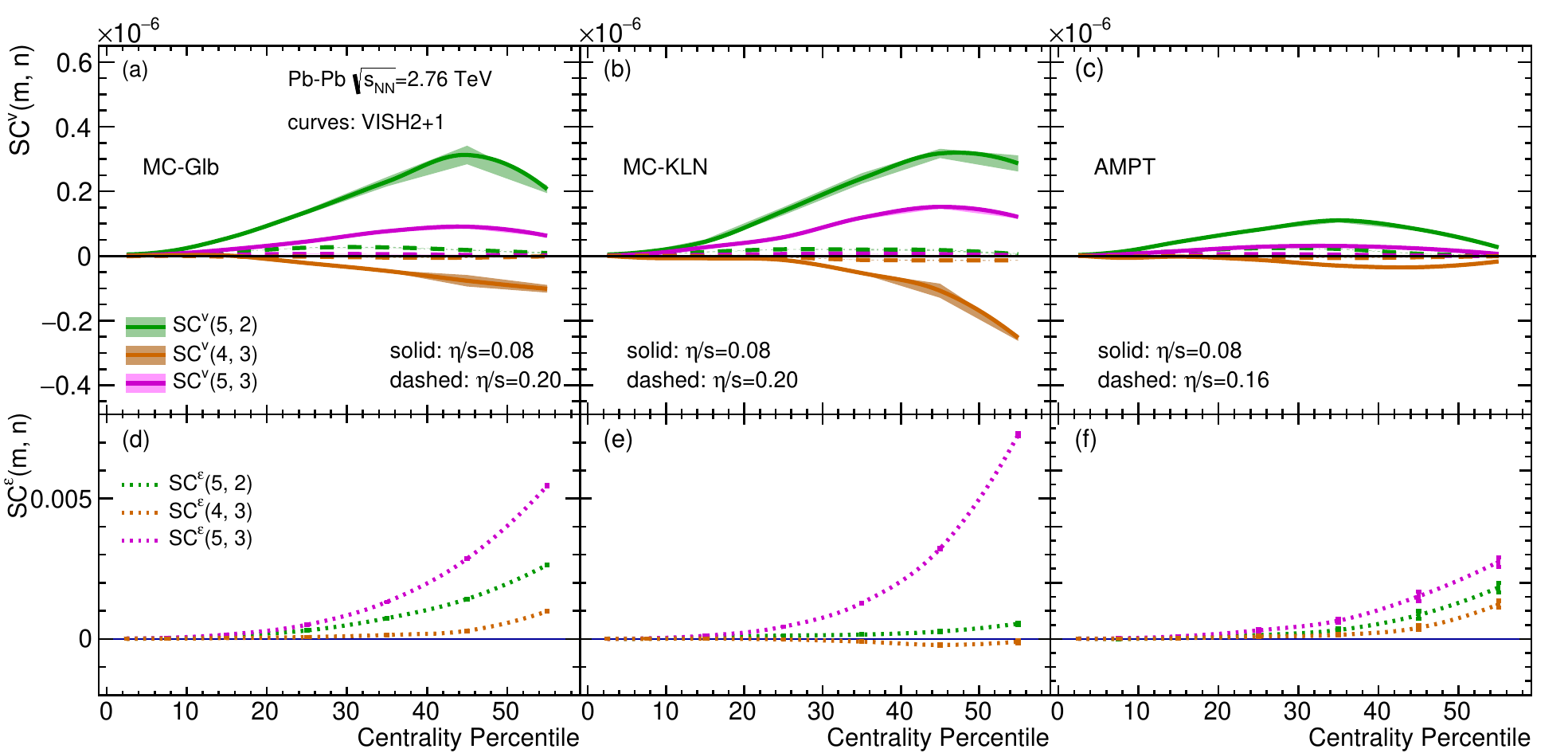}
  \caption{(Color online) Similar to Fig.~2, but for the predicted symmetric cumulants $SC^{v}(5, 2)$, $SC^{v}(5, 3)$ and $SC^{v}(4, 3)$ in 2.76 A TeV Pb--Pb collisions, together with the calculated symmetric cumulants of the initial eccentricity coefficients $SC^{\varepsilon}(5, 2)$, $SC^{v}(5, 3)$ and $SC^{\varepsilon}(4, 3)$ for {\tt MC-Glauber}, {\tt MC-KLN}, and {\tt AMPT} initial conditions, respectively.
 \label{fig:scvmvn2}}
\end{figure*}

\section{Results and discussion\label{sec:results}}
Before investigating the correlations between different flow harmonics, we firstly calculate the $\pt$-integrated flow $v_{2}$, $v_{3}$, and $v_{4}$ of all charged hadrons in 2.76A TeV Pb--Pb collisions, using event-by-event viscous hydrodynamics {\tt VISH2+1} with different combinations of initial conditions and specific shear viscosity. The comparison with the ALICE data~\cite{ALICE:2011ab} are shown in Fig.~\ref{fig:v2v3v4}. It demonstrates that, for the viscous hydrodynamic simulations with a uniform $\eta/s$, neither {\tt MC-Glauber} nor {\tt MC-KLN} initial conditions can simultaneously describe $v_{2}$, $v_{3}$, and $v_{4}$, as once reported in~\cite{Qiu:2011hf}. More specifically, for {\tt MC-Glauber} initial conditions, {\tt VISH2+1} with $\eta/s=0.08$ could nicely fit the integrated $v_{2}$ from central to semi-peripheral collisions, but overestimates $v_{3}$ and $v_{4}$ for the same centrality classes. For {\tt MC-KLN} initial conditions, {\tt VISH2+1} with $\eta/s=0.20$ reproduces the integrated flow $v_{2}$ but underestimates $v_{3}$ and $v_{4}$ for the presented centrality classes. Compared with these two results, hydrodynamic calculations with {\tt AMPT} initial conditions improves the descriptions of $v_n$ ($n=$ 2, 3, 4) with an additional smearing factor $\sigma$ during the initial energy depositions~\cite{Xu:2016hmp}.
Panel (c) shows that {\tt VISH2+1} with {\tt AMPT} initial condition and $\eta/s=0.08$ roughly describe $v_{2}$, $v_{3}$ and $v_{4}$ from central to semi-peripheral collisions~\footnote{Ref.~\cite{Bhalerao:2015iya} showed better descriptions of $v_n$ ($n=2, 3, 4$) for the {\tt VISH2+1} simulations with {\tt AMPT} initial conditions, especially for the centrality-dependent $v_2$. Compared with our calculations, which define the centrality bins through the distributions of initial total entropies, their centrality bins are cut by the empirical formula of {\tt AMPT} $c = \pi b^{2}/\sigma$~\cite{Xu:2011fi,Lin:2004en}. Since this paper is not aim to study the properties of flow harmonics, quantitatively, we continue to use the early parameters sets of {\tt AMPT} as used in ~\cite{Bhalerao:2015iya, Xu:2011fi,Lin:2004en}, rather than fine-tune them to obtain a better description of the centrality dependent $v_2$.}. Note that, for {\tt AMPT} initial conditions, we do not finely tune $\eta/s$ and other related parameters to obtain the best fit of $v_{n}$ (n=2, 3, 4), but continue to use one of the ``standard" specific shear viscosity $\eta/s=0.08$ as used for {\tt MC-Glauber} and {\tt MC-KLN} initial conditions.

\begin{figure*}[ht]
 \includegraphics[width=0.8\linewidth]{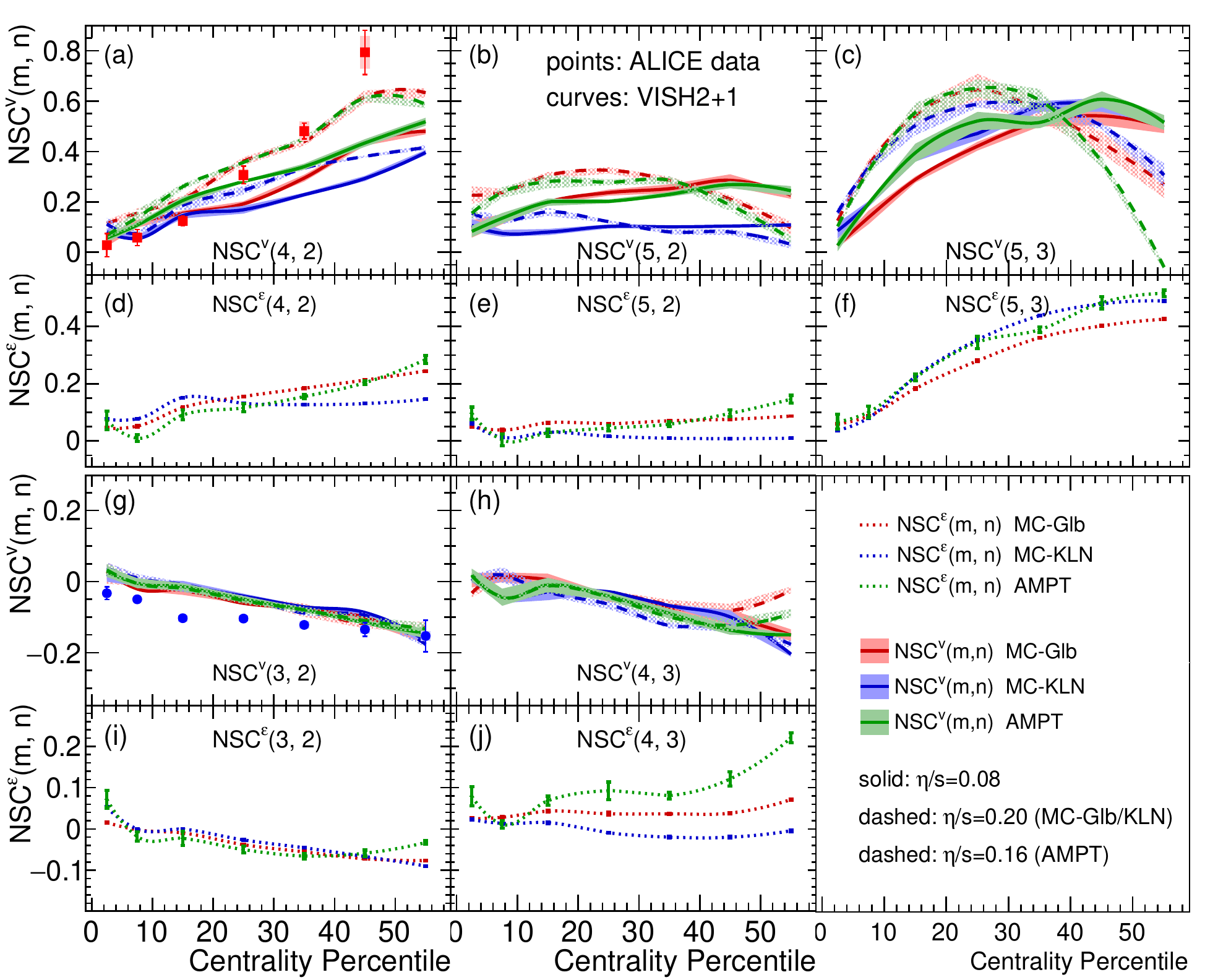}
  \caption{(Color online) Normalized symmetric cumulants $NSC^{v}(m, n)$ and normalized symmetric cumulants of the initial eccentricity coefficients $NSC^{\varepsilon}(m, n)$ in 2.76A TeV Pb--Pb collisions.
\label{fig:scRvmvn}}
\end{figure*}

\begin{figure*}[ht]
 \includegraphics[width=0.8\linewidth]{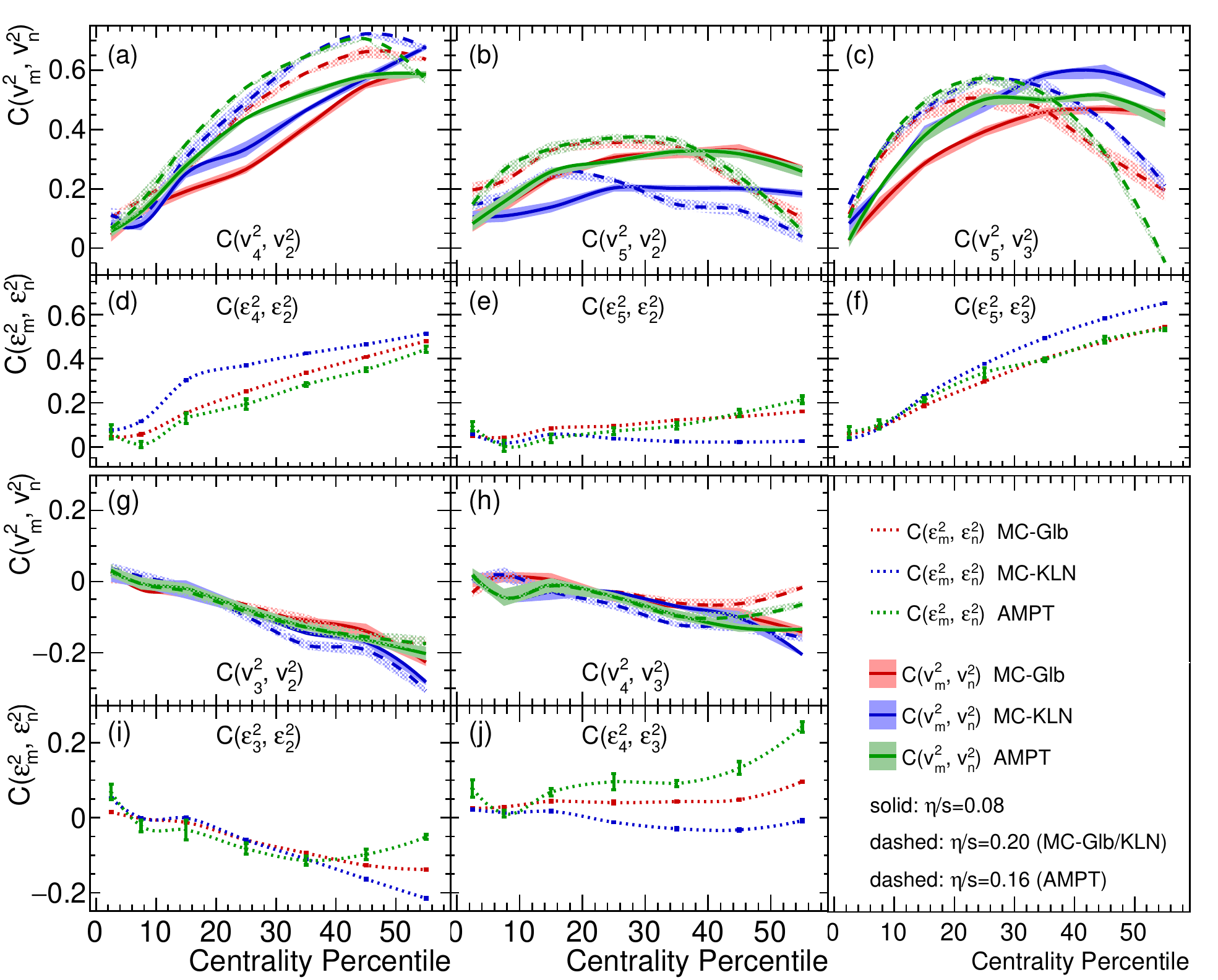}
  \caption{(Color online) The Pearson correlation coefficients of flow harmonics $C(v_{m}^{2}, v_{n}^{2})$ and the Pearson correlation coefficients of the initial eccentricity coefficients $C(\varepsilon_{m}^{2}, \varepsilon_{n}^{2})$ in 2.76A TeV Pb--Pb collisions.
 \label{fig:scCRvmvn}}
\end{figure*}

Using the same inputs and parameter sets, we calculate the symmetric cumulants $SC^{v}(m, n)$ in 2.76A TeV Pb--Pb collisions with {\tt VISH2+1}. The upper panels of Fig.~\ref{fig:scvmvn1} show the comparisons between our model calculations and the ALICE measurements. For all three initial conditions and different values of $\eta/s$, {\tt VISH2+1} could reproduce the typical features of the correlations between different flow harmonics, which shows negative values of $SC^{v}(3, 2)$ and positive values of $SC^{v}(4, 2)$. Therefore, both the experimental data and hydrodynamic calculations suggest that $v_2$ and $v_3$ are anti-correlated, while $v_2$ and $v_4$ are correlated. It also indicates that, for a specific event with larger value of $v_{2}$ above the event sample averaged $\langle v_{2} \rangle$, the probability of finding smaller value of $v_{3}$ below $\langle v_{3} \rangle$ and the probability of finding larger value of $v_{4}$ above $\langle v_{4} \rangle$ are both enhanced.
Fig.~\ref{fig:scvmvn1} also demonstrates that $SC^{v}(3, 2)$ and $SC^{v}(4, 2)$ are sensitive to the specific shear viscosity $\eta/s$ of the expanding fireball. For these three initial conditions, the anti-correlations between $v_{2}$ and $v_{3}$ and the correlations between $v_{2}$ and $v_{4}$ are both suppressed by the larger value of $\eta/s$.
This is due to the fact that the correlation strength of $SC^{v}(m, n)$ depends on the magnitudes of $v_{m}$ and $v_{n}$. Larger $\eta/s$ leads to a larger suppression of the flow harmonics $v_{n}$, which results in smaller values of $SC^{v}(m, n)$. This agrees with the conclusion from the transport model calculations~\cite{Bilandzic:2013kga}, which shows that stronger (anti-)correlations of $SC^{v}(m, n)$ are produced when using larger partonic cross sections (corresponding to smaller $\eta/s$~\cite{Xu:2011fe}).

In Fig.~\ref{fig:scvmvn1} (d), (e) and (f), we plot the symmetric cumulants of the initial eccentricity coefficients $SC^{\varepsilon}(3, 2)$ and $SC^{\varepsilon}(4, 2)$. For each initial condition, $SC^{\varepsilon}(3, 2)$ shows negative values and $SC^{\varepsilon}(4, 2)$ shows positive values, which demonstrates that $\varepsilon_2$ and $\varepsilon_3$ are anti-correlated, and $\varepsilon_2$ and $\varepsilon_4$ are correlated. The upper and lower panels of Fig.~\ref{fig:scvmvn1} also reveal that, although the signs of $SC^{v}(3, 2)$ and $SC^{v}(4, 2)$ are the same as the signs of $SC^{\varepsilon}(3, 2)$ and $SC^{\varepsilon}(4, 2)$, respectively, the correlation strength of them are strongly influenced by the viscous corrections of the QGP fireball. In peripheral collisions, both $SC^{\varepsilon}(3, 2)$ and $SC^{\varepsilon}(4, 2)$ shows larger correlation strength for these three initial conditions. However, the viscous fluid expansion has limited power to effectively convert the initial state correlations into the final state ones due to the largely reduced evolution time.

Although none of the above combination of  initial conditions and $\eta/s$ can quantitatively describes the data from ALICE, it is still impressive that event-by-event hydrodynamic simulations can correctly capture the sign of $SC^{v}(3, 2)$, $SC^{v}(4, 2)$ and roughly describe the centrality dependence.
Ref.~\cite{Niemi:2015qia} also showed that, although the EKRT+viscous hydrodynamic calculations with a temperature dependent $\eta/s(T)$ can nicely describe the centrality dependent integrated flow $v_{2}$, $v_{3}$ and $v_{4}$, the related model calculations can not quantitatively reproduce the centrality dependent $SC^{v}(3, 2)$ and $SC^{v}(4, 2)$ measurements. Meanwhile, the calculation from {\tt HIJING} simulations shows almost zero values of $SC^{v}(3, 2)$ and $SC^{v}(4, 2)$~\cite{ALICE:2016kpq}. These different theoretical calculations suggest that the correlations between different flow harmonics, i.e. $SC^{v}(3, 2)$ and $SC^{v}(4, 2)$, reflect the hydrodynamics response of the initial state correlations, which are more sensitive to the details of theoretical models than the individual $v_{n}$ coefficients alone.

In Fig.~\ref{fig:scvmvn2}, we predict the centrality dependent $SC^{v}(m, n)$ for other combinations of flow harmonics ($(m, n)=$ (5, 2), (5, 3) and (4, 3)), together with the calculations of $SC^{\varepsilon}(m, n)$ for the corresponding initial eccentricity coefficient pairs. For all three initial conditions, $SC^{v}(5, 2)$ and $SC^{v}(5, 3)$ are positive, and $SC^{v}(4, 3)$ are negative, which reveals that $v_{2}$ and $v_{5}$, $v_{3}$ and $v_{5}$ are correlated, while $v_{3}$ and $v_{4}$ are anti-correlated. Similar to $SC^{v}(3, 2)$ and $SC^{v}(4, 2)$, the magnitudes of $SC^{v}(5, 2)$, $SC^{v}(5, 3)$ and $SC^{v}(4, 3)$ are sensitive to the specific shear viscosity of QGP. Their correlation strengths become weaker with the increase of $\eta/s$. We also observe that the signs of $SC^{v}(5, 2)$ and $SC^{v}(5, 3)$ are the same as the corresponding initial state correlators $SC^{\varepsilon}(5, 2)$ and $SC^{\varepsilon}(5, 3)$. While, $SC^{v}(4, 3)$ and $SC^{\varepsilon}(4, 3)$ present opposite signs for {\tt MC-Glauber} and {\tt AMPT} initial conditions. In Refs.~\cite{Gardim:2011xv,Teaney:2012ke,Teaney:2013dta}, it has been found that the $v_4$ signals are influenced by both $\varepsilon_4$ and $\varepsilon_2^2$ of the initial conditions, where the $\varepsilon_2^2$ term makes the dominant contributions in non-central collisions~\cite{Yan:2015jma}. As a result, the anti-correlation between $\varepsilon_2$ and $\varepsilon_3$ significantly contributes to $SC^{v}(4, 3)$, leading to a changing sign of $SC^{v}(4, 3)$, when compared with $SC^{\varepsilon}(4, 3)$ for {\tt MC-Glauber} and {\tt AMPT} initial conditions.

\begin{figure*}[htpb]
\includegraphics[width=0.8\linewidth]{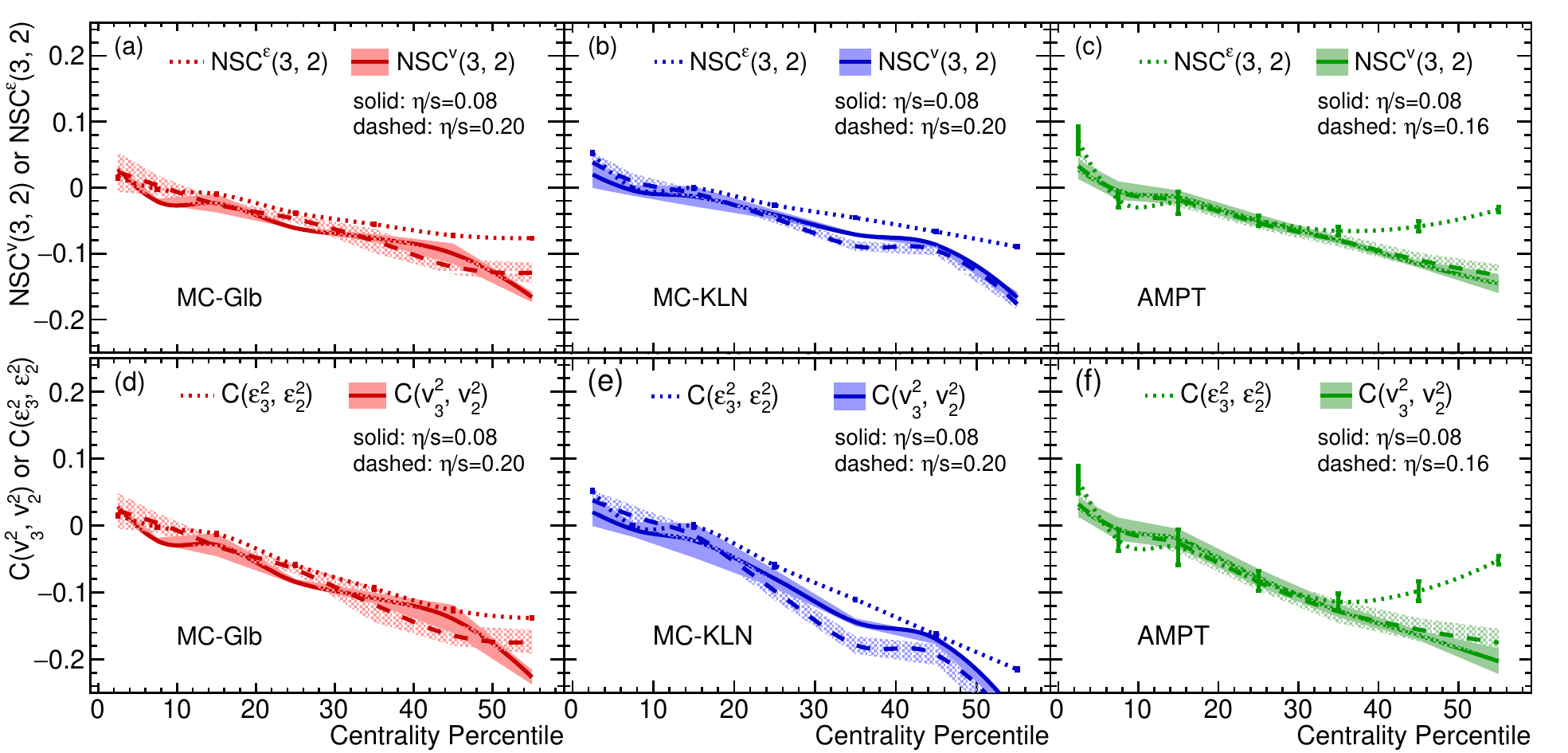}
  \caption{(Color online) Up panels: a comparison of the normalized symmetric cumulants $NSC^{v}(3, 2)$ and the normalized symmetric cumulants
  of the initial eccentricity coefficients $NSC^{\varepsilon}(3, 2)$ for {\tt MC-Glauber} (left), {\tt MC-KLN} (middle), and {\tt AMPT} (right) initial conditions. Lower panels: a similar comparison of the corresponding Pearson correlation coefficients $C(v_{3}^{2}, v_{2}^{2})$ and $C(\varepsilon_{3}^{2}, \varepsilon_{2}^{2})$.
   \label{fig:v3v2}}
\end{figure*}

In order to further study the correlations between flow harmonics, we normalize $SC^{v}(m, n)$ and $SC^{\varepsilon}(m, n)$ with $\langle v_{m}^2 \rangle\langle v_{n}^2 \rangle$ and $\langle \varepsilon_{m}^2 \rangle\langle \varepsilon_{n}^2 \rangle$ according to Eqs.~(\ref{eq:NSCvmvn}) and~(\ref{eq:NSCemen}), respectively. The normalized symmetric cumulants $NSC^{v}(m, n)$ and the corresponding initial state normalized correlators $NSC^{\varepsilon}(m, n)$ are plotted in Fig.~\ref{fig:scRvmvn}, where panel (a) and (g) also show the corresponding measurements from ALICE~\cite{ALICE:2016kpq}. We find $NSC^{v}(4, 2)$, $NSC^{v}(5, 2)$, and $NSC^{v}(5, 3)$ are sensitive to both initial conditions and the specific shear viscosity $\eta/s$. More specifically, they show sizable changes to the change of $\eta/s$ for certain initial condition. Meanwhile, they are also influenced by the initial conditions. Their corresponding $NSC^{\varepsilon}$ correlators also separate in different initial conditions. For {\tt AMPT} and {\tt MC-Glauber} initial conditions, the {\tt VISH2+1} calculations roughly fit $NSC^{v}(4, 2)$ data with $\eta/s=0.16$ and $\eta/s=0.2$, which demonstrates that such normalized symmetric cumulants can help to constrain the QGP viscosity. We also find, for these three investigated initial conditions,  $NSC^{\varepsilon}(3, 2)$ are almost overlap from central to semi-central collisions which only slightly split in peripheral collisions. At the same time, the normalized symmetric cumulants $NSC^{v}(3, 2)$ is insensitive to the QGP shear viscosity since $v_2$ and $v_3$ are roughly proportional to $\varepsilon_2$ and $\varepsilon_3$. As a result, these different $NSC^{v}(3, 2)$ curves in Fig.~4 (g) are almost overlap with each other, which also roughly fit the normalized ALICE data.
Similar to the case of $SC^{v}(4, 3)$ and $SC^{\varepsilon}(4, 3)$ in Fig.~\ref{fig:scvmvn2}, $NSC^{v}(4, 3)$ does not follow the sign of $NSC^{\varepsilon}(4, 3)$ for both {\tt MC-Glauber} and {\tt AMPT} initial conditions due to the nonlinear and dominant contribution of $\varepsilon_{2}^{2}$ to $v_4$ from semi-central to peripheral collisions. Panel (h) and (j) also show that, although $NSC^{\varepsilon}(4, 3)$ is strongly depends on the initial conditions, $NSC^{v}(4, 3)$ is not very sensitive to the initial conditions, which roughly overlaps from central to semi-peripheral collisions for different $\eta/s$. In a recent work, the $NSC^{v}(m,n)$ are expressed in terms of event-plane correlations and moments of $v_{2}$ and $v_{3}$ (see Eqs.~(8) and (9) in~\cite{Giacalone:2016afq}). Considering the relative flow fluctuations of $v_{3}$ is stronger than $v_{2}$, one expects that $\left< v_{2}^{4} \right>/ \left< v_{2}^{2} \right>^{2}$ is smaller than $\left< v_{3}^{4} \right> / \left< v_{3}^{2} \right>^{2}$~\cite{Bhalerao:2014xra}, which in the end gives smaller values for $NSC^{v}(5,2)$ than $NSC^{v}(5,3)$. This is indeed observed in Fig.~\ref{fig:scRvmvn} (b) and (c).
On the other hand, it was predicted the $NSC^{v}(m,n)$ correlators that involve $v_{4}$ or $v_{5}$ increase with $\eta/s$ in the same way as the event-plane correlations~\cite{Teaney:2012ke, Teaney:2012gu}. This seems agree with what shown in panel (a) as well as the cases from central to semi-peripheral collisions in pannel (b) and (c) of Fig.~\ref{fig:scRvmvn}, but \underline{}in contrast to the results for peripheral collisions in panel (b) and (c). Examinations on the Eqs.~(8) and (9) in~\cite{Giacalone:2016afq} and the corresponding assumption used in nonlinear hydrodynamic response phenomena with future experimental data are necessary to explain the difference between our results and those theoretical predictions in~\cite{Giacalone:2016afq}.

Besides $NSC^{v}(m, n)$, one can also investigate the correlations between different flow harmonics through the Pearson correlation coefficients $C(v_{m}^{2}, v_{n}^{2})$ defined by Eq.~(\ref{eq:SCCRvmvn}). As introduced in Sec.~\ref{sec:setup}, $C(v_{m}^{2}, v_{n}^{2})$ could further evaluate the linear relationship between $v_m$ and $v_n$. In Fig.~\ref{fig:scCRvmvn}, we plot the centrality dependent $C(v_{m}^{2}, v_{n}^{2})$ calculated from the {\tt VISH2+1} with different initial conditions and $\eta/s$, together with the corresponding initial state correlators $C(\varepsilon_{m}^{2}, \varepsilon_{n}^{2})$. We find that the absolute values of all $C(v_{m}^{2}, v_{n}^{2})$ and $C(\varepsilon_{m}^{2}, \varepsilon_{n}^{2})$ do not equal to 1, which indicate none of the ($v_{m}^{2}, v_{n}^{2}$), ($\varepsilon_{m}^{2}, \varepsilon_{n}^{2}$) pairs are linearly correlated or anti-correlated. For different $(m, n)$ pairs, $C(v_{m}^{2}, v_{n}^{2})$ shows similar dependence on initial conditions and the specific shear viscosity $\eta/s$ as the case of $NSC^{v}(m, n)$. More specifically,  $C(v_{4}^{2}, v_{2}^{2})$, $C(v_{5}^{2}, v_{2}^{2})$, and $C(v_{5}^{2}, v_{3}^{2})$ strongly depend on both initial conditions and $\eta/s$, however, $C(v_{3}^{2}, v_{2}^{2})$ is insensitive to initial conditions and $\eta/s$. Meanwhile, the corresponding $C(\varepsilon_{3}^{2}, \varepsilon_{2}^{2})$ from different initial conditions almost overlap with each other from central to semi-peripheral collisions. Although $C(\varepsilon_{4}^{2}, \varepsilon_{3}^{2})$ from different initial conditions show significant separations, $C(v_{4}^{2}, v_{3}^{2})$ from our model calculations is not very sensitive to initial conditions since $v_4$ is largely influence by $\varepsilon^2_2$ from semi-central to peripheral collisions.

\begin{figure*}[ht]
 \includegraphics[width=0.8\linewidth]{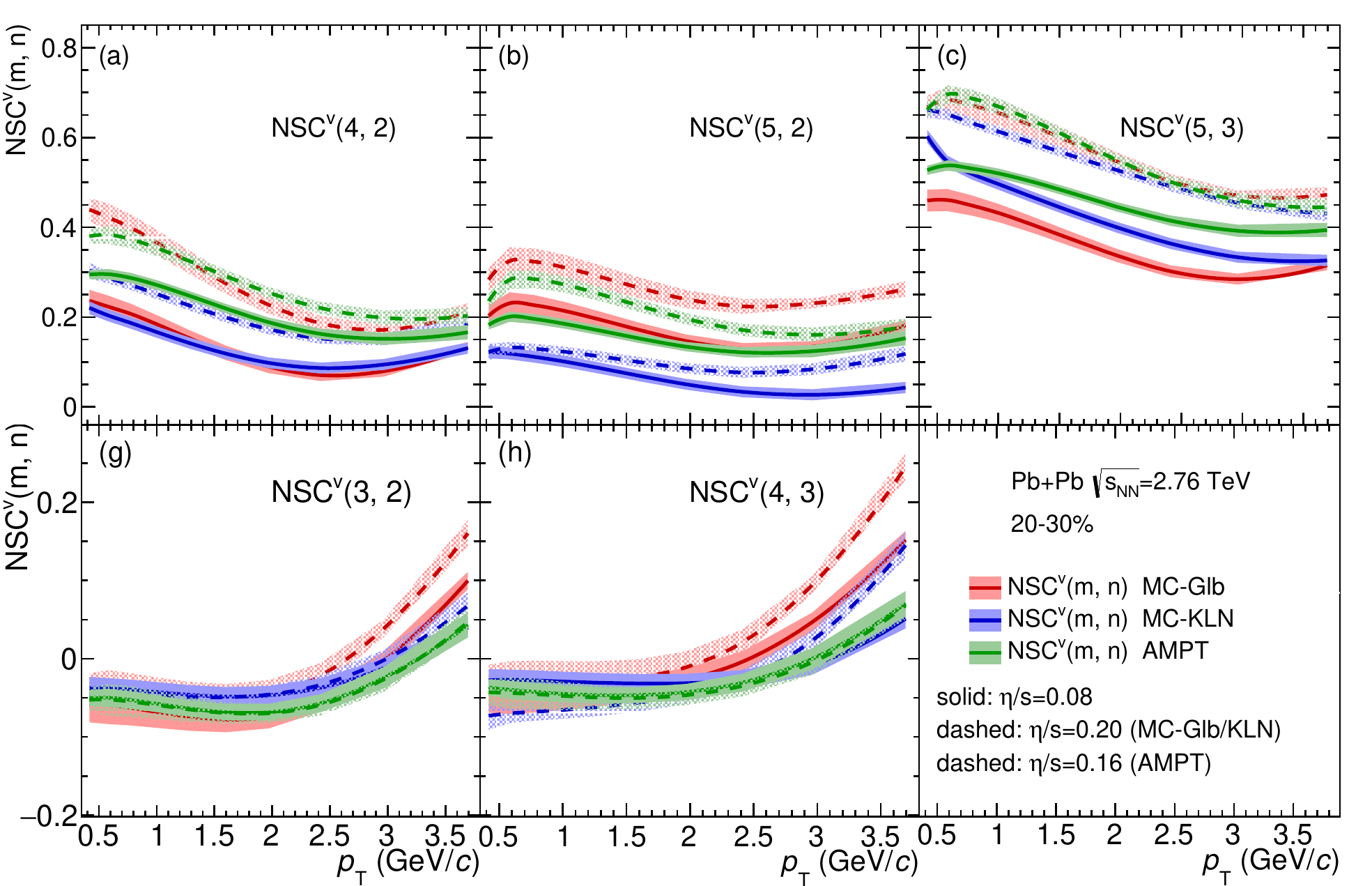}
  \caption{(Color online) $\pt$ dependent normalized symmetric cumulants $NSC^{v}(m, n)$ at 20-30\% centrality in 2.76 A TeV Pb--Pb collisions, calculated from {\tt VISH2+1} with {\tt MC-Glauber} (red), {\tt MC-KLN} (blue), and {\tt AMPT} (green) initial conditions.
\label{fig:NSCpt2030}}
\end{figure*}

Both Fig.~\ref{fig:scRvmvn} and Fig.~\ref{fig:scCRvmvn} demonstrate that, from central to semi-peripheral collisions, the normalized symmetric cumulant $NSC^{v}(3, 2)$ and Pearson correlation coefficient $C(v_{3}^{2}, v_{2}^{2})$ are insensitive to the specific shear viscosity $\eta/s$  for various initial conditions. In Fig.~\ref{fig:v3v2}, we directly compare $NSC^{v}(3, 2)$ and $C(v_{3}^{2}, v_{2}^{2})$ with the corresponding initial state correlators $NSC^{\varepsilon}(3, 2)$ and $C(\varepsilon_{3}^{2}, \varepsilon_{2}^{2})$, respectively. We observe for each initial condition, $NSC^{v}(3, 2)$ are almost overlap with the $NSC^{\varepsilon}(3, 2)$ within the statistical uncertainties from central to semi-central collisions, despite of the input $\eta/s$. Similarly, $C(v_{3}^{2}, v_{2}^{2})$ and $C(\varepsilon_{3}^{2}, \varepsilon_{2}^{2})$ also roughly overlap from central to semi-central collisions. These results demonstrate that the $NSC^{v}(3, 2)$  and $C(v_{3}^{2}, v_{2}^{2})$ are mainly determined by corresponding correlators $NSC^{\varepsilon}(3, 2)$ and $C(\varepsilon_{3}^{2}, \varepsilon_{2}^{2})$ from the initial state.

Figure~\ref{fig:NSCpt2030} presents $NSC^{v}(m, n)$ as a function of $\pt$ in 20-30\% Pb--Pb collisions. Besides different sensitivities to initial conditions and $\eta/s$ as discussed above, we notice that $NSC^{v}(m, n)$ also shows different $\pt$ dependence. More specifically, $NSC^{v}(4, 2)$, $NSC^{v}(5, 2)$ and $NSC^{v}(5, 3)$ are positive for the entire $\pt$ range, while $NSC^{v}(3, 2)$ and $NSC^{v}(4, 3)$ are negative at low $\pt$ but change to positive for $\pt >3$ GeV/$c$. The trend is qualitatively agreed with the conclusion obtained with linear correlation function $c(v_{m}, v_{n})$ reported in~\cite{Niemi:2012aj}, although different moments of $v_{m}$ and $v_n$ are used. Future investigations on non-linear hydrodynamic response in higher flow harmonics will help us to better understand the observed different behaviors of $\pt$ dependent $NSC^{v}(m, n)$.

\section{Summary\label{sec:summary}}
In this paper, we investigate the correlations of flow harmonics in 2.76A TeV Pb--Pb collisions using the event-by-event viscous hydrodynamics {\tt VISH2+1} with {\tt MC-Glauber}, {\tt MC-KLN}, and {\tt AMPT} initial conditions. We found the symmetric cumulants $SC^{v}(m, n)$ are sensitive to both initial conditions and the specific shear viscosity $\eta/s$. When compared to the ALICE data, our {\tt VISH2+1} calculations could qualitatively describe $SC^{v}(3, 2)$ and $SC^{v}(4, 2)$ for different initial conditions, which demonstrate that $v_2$ and $v_4$ are correlated and $v_2$ and $v_3$ are anti-correlated. We also predicted other symmetric cumulants with different $(m, n)$ combinations and found  $v_2$ and $v_5$, $v_3$ and $v_5$ are correlated, $v_3$ and $v_4$ are anti-correlated at various centralities.

In addition, we investigate the normalized symmetric cumulants $NSC^{v}(m, n)$ and the Pearson correlation coefficients $C(v_{m}^{2}, v_{n}^{2})$. We found $NSC^{v}(3, 2)$ and $C(v_{3}^{2}, v_{2}^{2})$ are mainly determined by corresponding $NSC^{\varepsilon}(3, 2)$ and $C(\varepsilon_{3}^{2}, \varepsilon_{2}^{2})$ correlators from the initial state, which roughly overlap from central to semi-peripherial collisions for the three initial conditions used in our calculations. Furthermore, $NSC^{v}(3, 2)$ and $C(v_{3}^{2}, v_{2}^{2})$ are insensitive to the specific shear viscosity $\eta/s$ in the hydrodynamic simulations, since both $v_2$ and $v_3$ are approximately linearly response to $\varepsilon_2$ and $\varepsilon_3$ of the initial state. In contrast, $NSC^{v}(4, 2)$, $NSC^{v}(5, 2)$, $NSC^{v}(5, 3)$, as well as $C(v_{4}^{2}, v_{2}^{2})$, $C(v_{5}^{2}, v_{2}^{2})$, and $C(v_{5}^{2}, v_{3}^{2})$, are sensitive to both initial conditions and $\eta/s$. We also found, for both {\tt MC-Glauber} and {\tt AMPT} initial conditions, $NSC^{v}(4, 3)$ does not follow the sign of $NSC^{\varepsilon}(4, 3)$ due to the nonlinear mode couplings between $v_4$ and $v_2$. Correspondingly, although $NSC^{\varepsilon}(4, 3)$ is strongly depends on the initial conditions, $NSC^{v}(4, 3)$ curves are not very sensitive to the initial conditions, which are also roughly overlap for different initial conditions and specific shear viscosity $\eta/s$ from central to semi-peripheral collisions.
We also studied the $\pt$ dependent normalized symmetric cumulants and observed that $NSC^{v}(4, 2)$, $NSC^{v}(5, 2)$ and $NSC^{v}(5, 3)$ are positive while $NSC^{v}(3, 2)$ and $NSC^{v}(4, 3)$ change sign from negative to positive at $\pt \sim$ 3 GeV/$c$.

Compared to individual $v_{n}$ coefficients, the correlations between different flow harmonics are more sensitive to the details of theoretical calculations. Future experimental measurements on the predicted observables, including symmetric cumulants  $SC^{v}(5, 2)$,  $SC^{v}(5, 3)$, $SC^{v}(4, 3)$ and the normalized symmetric cumulants $NSC(m, n)$, the Pearson correlation coefficients $C(v_{m}^{2}, v_{n}^{2})$, and the further related hydrodynamic investigations will shed new light into the nature of the initial state fluctuations and the properties of the QGP fireball created in the ultra-relativistic heavy ion collisions.

\acknowledgments
We thank J.~J.~Gaardh\o je, U.~Heinz, M.~Luzum, J.~Noronha-Hostler, J.~Y.~Ollitrault and L.~Yan for helpful discussions.
This work is supported by the NSFC and the MOST under grant Nos.11435001 and 2015CB856900, and partially supported by China Postdoctoral Science Foundation under grant No. 2015M570878 and 2015M580908, by the Danish Council for Independent Research, Natural Sciences, and the Danish National Research Foundation (Danmarks Grundforskningsfond). We gratefully acknowledge the extensive computing resources provided to us by Super-computing Center of Chinese Academy of Science (SCCAS) and Tianhe-1A from the National Supercomputing Center in Tianjin, China.


\begin{thebibliography}{99}
\bibitem{Rev-Arsene:2004fa}
  I.~Arsene {\it et al.} (BRAHMS Collaboration),
  Nucl.\ Phys.\  {\bf A757}, 1 (2005);
  B.~B.~Back {\it et al.} (PHOBOS Collaboration),
  {\it ibid.,} p.\,28;
  J.~Adams {\it et al.} (STAR Collaboration),
  {\it ibid.,} p.\,102;
  K.~Adcox {\it et al.} (PHENIX Collaboration),
  {\it ibid.,} p.\,184.

\bibitem{Gyulassy:2004vg}
M.~Gyulassy,
in {\it Structure and dynamics of elementary matter}, edited by
W. Greiner {\it et al.}, NATO science series II: Mathematics, physics
and chemistry, Vol. 166 (Kluwer Academic, Dordrecht, 2004), p.~159-182
  M.~Gyulassy and L.~McLerran,
  Nucl.\ Phys.\  {\bf A750}, 30 (2005);
  E.~V.~Shuryak,
  {\it ibid.}, p.\,64.

\bibitem{Muller:2006ee}
  B.~Muller and J.~L.~Nagle,
  Ann.\ Rev.\ Nucl.\ Part.\ Sci.\  {\bf 56}, 93 (2006).

\bibitem{Muller:2012zq}
  B.~Muller, J.~Schukraft and B.~Wyslouch,
  Ann.\ Rev.\ Nucl.\ Part.\ Sci.\  {\bf 62}, 361 (2012).

\bibitem{Ollitrault:1992bk}
  J.~-Y.~Ollitrault,
  Phys.\ Rev.\ D {\bf 46}, 229 (1992).

  \bibitem{Voloshin:2008dg}
  S.~A.~Voloshin, A.~M.~Poskanzer and R.~Snellings,
  arXiv:0809.2949 [nucl-ex].

\bibitem{Luzum:2013yya}
  M.~Luzum and H.~Petersen,
  J.\ Phys.\ G {\bf 41}, 063102 (2014).

\bibitem{Heinz:2013th}
 U.~Heinz and R.~Snellings,
 Annu.\  Rev.\  Nucl.\  Part.\  Sci.\  {\bf 63}, 123 (2013).

\bibitem{Gale:2013da}
 C.~Gale, S.~Jeon and B.~Schenke,
 Int.\ J.\ Mod.\ Phys.\ A {\bf 28}, 1340011 (2013).

\bibitem{ALICE:2011ab}
  K.~Aamodt {\it et al.} [ALICE Collaboration],
  Phys.\ Rev.\ Lett.\  {\bf 107}, 032301 (2011).

\bibitem{Chatrchyan:2013kba}
  S.~Chatrchyan {\it et al.} [CMS Collaboration],
  Phys.\ Rev.\ C {\bf 89}, no. 4, 044906 (2014).

\bibitem{Qiu:2011iv}
 Z.~Qiu and U.~Heinz,
 Phys.\ Rev.\ C {\bf 84}, 024911 (2011).

\bibitem{Petersen:2010cw}
 H.~Petersen, G.~Y.~Qin, S.~A.~Bass and B.~Muller,
 Phys.\ Rev.\  C {\bf 82}, 041901 (2010);
 G.~Y.~Qin, H.~Petersen, S.~A.~Bass and B.~Muller,
 Phys.\ Rev.\  C {\bf 82}, 064903 (2010).

\bibitem{Holopainen:2010gz}
 H.~Holopainen, H.~Niemi and K.~J.~Eskola,
 Phys.\ Rev.\  C {\bf 83}, 034901 (2011).

\bibitem{Song:2010mg}
  H.~Song, S.~A.~Bass, U.~Heinz, T.~Hirano and C.~Shen,
  Phys.\ Rev.\ Lett.\  {\bf 106}, 192301 (2011);
 Phys.\ Rev.\ C {\bf 83}, 054910 (2011).

\bibitem{Song:2013qma}
  H.~Song, S.~Bass and U.~W.~Heinz,
  Phys.\ Rev.\ C {\bf 89}, no. 3, 034919 (2014);
   X.~Zhu, F.~Meng, H.~Song and Y.~X.~Liu,
   Phys.\ Rev.\ C {\bf 91}, no. 3, 034904 (2015).


\bibitem{Song:2012ua}
 H.~Song,
 Nucl.\ Phys.\ A {\bf 904-905}, 114c (2013);
 arXiv:1210.5778 [nucl-th];
 Pramana {\bf 84}, 703 (2015).

\bibitem{Schenke:2010rr}
 B.~Schenke, S.~Jeon and C.~Gale,
 Phys.\ Rev.\ Lett.\  {\bf 106}, 042301 (2011);
 Phys.\ Rev.\  C {\bf 85}, 024901 (2012).

\bibitem{Gale:2012rq}
 C.~Gale, S.~Jeon, B.~Schenke, P.~Tribedy and R.~Venugopalan,
 Phys.\ Rev.\ Lett.\  {\bf 110}, 012302 (2013).

\bibitem{Xu:2016hmp}
  H.~j.~Xu, Z.~Li and H.~Song,
  Phys.\ Rev.\ C {\bf 93}, no. 6, 064905 (2016).

\bibitem{Andronic:2000cx}
  A.~Andronic {\it et al.} [FOPI Collaboration],
  Nucl.\ Phys.\ A {\bf 679}, 765 (2001).

\bibitem{Chung:2001qr}
  P.~Chung {\it et al.} [E895 Collaboration],
  Phys.\ Rev.\ C {\bf 66}, 021901 (2002).


\bibitem{Adams:2003zg}
  J.~Adams {\it et al.} [STAR Collaboration],
  Phys.\ Rev.\ Lett.\  {\bf 92}, 062301 (2004).


\bibitem{Aad:2014fla}
  G.~Aad {\it et al.} [ATLAS Collaboration],
  Phys.\ Rev.\ C {\bf 90}, no. 2, 024905 (2014).

\bibitem{Qiu:2012uy}
  Z.~Qiu and U.~Heinz,
  Phys.\ Lett.\ B {\bf 717}, 261 (2012).

\bibitem{Teaney:2012gu}
  D.~Teaney and L.~Yan,
  Nucl.\ Phys.\ A {\bf 904-905}, 365c (2013).

 \bibitem{Jia:2012ju}
  J.~Jia and D.~Teaney,
  Eur.\ Phys.\ J.\ C {\bf 73}, 2558 (2013).

 \bibitem{Jia:2012ma}
  J.~Jia and S.~Mohapatra,
  Eur.\ Phys.\ J.\ C {\bf 73}, 2510 (2013).

\bibitem{Niemi:2015qia}
 H.~Niemi, K.~J.~Eskola and R.~Paatelainen,
 Phys.\ Rev.\ C {\bf 93}, no. 2, 024907 (2016).


\bibitem{Bilandzic:2013kga}
  A.~Bilandzic, C.~H.~Christensen, K.~Gulbrandsen, A.~Hansen and Y.~Zhou,
  Phys.\ Rev.\ C {\bf 89}, no. 6, 064904 (2014).

\bibitem{Bhalerao:2014xra}
  R.~S.~Bhalerao, J.~Y.~Ollitrault and S.~Pal,
  Phys.\ Lett.\ B {\bf 742}, 94 (2015).

\bibitem{Zhou:2015eya}
 Y.~Zhou, K.~Xiao, Z.~Feng, F.~Liu and R.~Snellings,
  Phys.\ Rev.\ C {\bf 93}, no. 3, 034909 (2016).

\bibitem{ALICE:2016kpq}
  J.~Adam {\it et al.} [ALICE Collaboration],
  arXiv:1604.07663 [nucl-ex];
  Y.~Zhou [ALICE Collaboration],
  arXiv:1512.05397 [nucl-ex].

\bibitem{Niemi:2012aj}
  H.~Niemi, G.~S.~Denicol, H.~Holopainen and P.~Huovinen,
  Phys.\ Rev.\ C {\bf 87}, no. 5, 054901 (2013).

\bibitem{Giacalone:2016afq}
  G.~Giacalone, L.~Yan, J.~Noronha-Hostler and J.~Y.~Ollitrault,
  arXiv:1605.08303 [nucl-th].

\bibitem{Qian:2016pau}
  J.~Qian and U.~Heinz,
  arXiv:1607.01732 [nucl-th].

\bibitem{Aad:2015lwa}
  G.~Aad {\it et al.} [ATLAS Collaboration],
  Phys.\ Rev.\ C {\bf 92}, no. 3, 034903 (2015).

\bibitem{Schukraft:2012ah}
  J.~Schukraft, A.~Timmins and S.~A.~Voloshin,
  Phys.\ Lett.\ B {\bf 719}, 394 (2013).

\bibitem{Song:2007fn}
  H.~Song and U.~Heinz,
  Phys.\ Lett.\  {\bf B658}, 279 (2008);
  Phys.\ Rev.\  C {\bf 77}, 064901 (2008);
  Phys.\ Rev.\ C {\bf 78}, 024902 (2008);
  H.~Song, Ph.D Thesis, The Ohio State University, August 2009,
  arXiv:0908.3656 [nucl-th].

\bibitem{Shen:2014vra}
  C.~Shen, Z.~Qiu, H.~Song, J.~Bernhard, S.~Bass and U.~Heinz,
  Comput.\ Phys.\ Commun.\  {\bf 199}, 61 (2016).

\bibitem{Qiu:2011hf}
  Z.~Qiu, C.~Shen and U.~Heinz,
  Phys.\ Lett.\ B {\bf 707}, 151 (2012).

\bibitem{Huovinen:2009yb}
  P.~Huovinen and P.~Petreczky,
  Nucl.\ Phys.\  {\bf A837}, 26 (2010).

\bibitem{Shen:2010uy}
  C.~Shen, U.~Heinz, P.~Huovinen and H.~Song,
  Phys.\ Rev.\ C {\bf 82}, 054904 (2010).

\bibitem{Cooper:1974mv}
  F.~Cooper and G.~Frye,
  Phys.\ Rev.\ D {\bf 10}, 186 (1974).

 \bibitem{Bhalerao:2015iya}
  R.~S.~Bhalerao, A.~Jaiswal and S.~Pal,
  Phys.\ Rev.\ C {\bf 92}, no. 1, 014903 (2015).

\bibitem{Pang:2012he}
  L.~Pang, Q.~Wang and X.~N.~Wang,
  Phys.\ Rev.\ C {\bf 86}, 024911 (2012).

\bibitem{Moreland:2014oya}
  J.~S.~Moreland, J.~E.~Bernhard and S.~A.~Bass,
  Phys.\ Rev.\ C {\bf 92}, no. 1, 011901 (2015).

\bibitem{Shen:2011eg}
  C.~Shen, U.~Heinz, P.~Huovinen and H.~Song,
  Phys.\ Rev.\ C {\bf 84}, 044903 (2011).

\bibitem{Kolb:2000sd}
  P.~F.~Kolb, J.~Sollfrank and U.~W.~Heinz,
  Phys.\ Rev.\ C {\bf 62}, 054909 (2000).

\bibitem{Kharzeev:2000ph}
  D.~Kharzeev and M.~Nardi,
  Phys.\ Lett.\ B {\bf 507}, 121 (2001);
  D.~Kharzeev, E.~Levin and M.~Nardi,
  Nucl.\ Phys.\ A {\bf 730}, 448 (2004); {\bf 743}, 329 (2004)]; {\bf 747}, 609 (2005).


\bibitem{Miller:2007ri}
  M.~L.~Miller, K.~Reygers, S.~J.~Sanders and P.~Steinberg,
  Ann.\ Rev.\ Nucl.\ Part.\ Sci.\  {\bf 57}, 205 (2007).

\bibitem{Drescher:2006ca}
  H.~J.~Drescher and Y.~Nara,
  Phys.\ Rev.\  C {\bf 75}, 034905 (2007);
   Phys.\ Rev.\  C {\bf 76}, 041903(R) (2007).
%
\bibitem{Hirano:2009ah}
  T.~Hirano and Y.~Nara,
  Phys.\ Rev.\  C {\bf 79}, 064904 (2009);
  T.~Hirano, P.~Huovinen and Y.~Nara,
  Phys.\ Rev.\ C {\bf 83}, 021902 (2011).

\bibitem{Oliinychenko:2015lva}
  D.~Oliinychenko and H.~Petersen,
  Phys.\ Rev.\ C {\bf 93}, no. 3, 034905 (2016).

\bibitem{Xu-Song}
  H. Xu and H. Song, unpublished note.

\bibitem{PCC}
K.~Pearson,
Proceedings of the Royal Society of London, {\bf 58}, 240-242, (1895);
A.~Ly, M.~Marsman, and E. Wagenmakers,
arXiv:1510.01188 [math.ST].


\bibitem{Xu:2011fi}
  J.~Xu and C.~M.~Ko,
  Phys.\ Rev.\ C {\bf 83}, 034904 (2011).

 \bibitem{Lin:2004en}
  Z.~W.~Lin, C.~M.~Ko, B.~A.~Li, B.~Zhang and S.~Pal,
  Phys.\ Rev.\ C {\bf 72}, 064901 (2005).

\bibitem{Xu:2011fe}
  J.~Xu and C.~M.~Ko,
  Phys.\ Rev.\ C {\bf 84}, 014903 (2011).


\bibitem{Gardim:2011xv}
    F.~G.~Gardim, F.~Grassi, M.~Luzum and J.~Y.~Ollitrault,
    Phys.\ Rev.\ C {\bf 85}, 024908 (2012).

\bibitem{Teaney:2012ke}
  D.~Teaney and L.~Yan,
  Phys.\ Rev.\ C {\bf 86}, 044908 (2012).
\bibitem{Teaney:2013dta}
  D.~Teaney and L.~Yan,
  Phys.\ Rev.\ C {\bf 90}, no. 2, 024902 (2014).

\bibitem{Yan:2015jma}
  L.~Yan and J.~Y.~Ollitrault,
  Phys.\ Lett.\ B {\bf 744}, 82 (2015).

\end{thebibliography}
\end{document}